\documentclass{article}
\usepackage{amssymb}

\usepackage[english,francais]{babel}

\newtheorem{theorem}{Theorem}[section]

\newtheorem{e-proposition}[theorem]{Proposition}

\newtheorem{e-definition}[theorem]{Definition\rm}

\font\bb=msbm10 at 12pt
\font\bbbis=msbm10 at 10pt
\def\og{\leavevmode\raise.3ex\hbox{$\scriptscriptstyle\langle\!\langle$~}}
\def\fg{\leavevmode\raise.3ex\hbox{~$\!\scriptscriptstyle\,\rangle\!\rangle$}}
\def\rR{\hbox{\bb R}}
\def\zZ{\hbox{\bb Z}}

\begin{document}
% place in the next line the header (rubrique) chosen for your article,
% if you know it (you can also have 2, format : Header1/Header2
%\centerline{}
% \author{Name\thanksref{label2}}
\selectlanguage{english}
\title{Concentration of the first eigenfunction for a second order elliptic operator}

\author{David Holcman \thanks{
Weizmann Institute of Science, department of Mathematics, Rehovot 76100,
Israel. D. H. is incumbent of the Madeleine Haas Russel Career Development
Chair.Visiting  Departement de Biologie, Ecole Normale Sup\'erieure, 
46, rue d'Ulm, 75005 Paris, France. } \and Ivan Kupka\thanks{
Universit\'e Paris VI, department of Mathematics, 175 rue du Chevaleret
75013 Paris, France. I.K. thanks the Weizmann Institute of Science for its
hospitality during the preparation of this paper.}}
\date{}
\maketitle

% The [label1] can be suppressed if there is only one address for all authors
% If you know the dates of reception, and acceptation you can put them now;
%  idem the name of the person presenting the Note
%\medskip
%\begin{center}
%{\small Received *****; accepted after revision +++++\\
%Presented by Thierry Aubin}
%\end{center}

\begin{abstract}
\selectlanguage{english}
We study the semi-classical limits of the first eigenfunction of a positive
second order operator on a compact Riemannian manifold when the diffusion constant
 $\epsilon$ goes to zero. We assume that the first order term is given by a vector
 field $b$, whose recurrent components  are either   hyperbolic points  or cycles 
or two dimensional torii. The limits of the normalized  eigenfunctions
concentrate on the recurrent sets of maximal dimension where the topological pressure \cite{Kifer90} 
is attained. On the cycles and torii, the limit measures are absolutely continuous 
with respect to the invariant probability measure on these sets. We have determined these limit 
measures, using a  blow-up analysis.

%{\it To cite this article: D. Holcman, I. Kupka, C. R. Acad. Sci. Paris, Ser. I 340 (2005).}

\vskip 0.5\baselineskip

\selectlanguage{francais}
% Text of abstract in French
\noindent{\bf R\'esum\'e} \vskip 0.5\baselineskip \noindent
{\bf Concentration de la premi\`ere fonction propre pour un op\'erateur elliptique du second ordre}
%Your resume in French here.

Nous \'etudions sur une vari\'et\'e Riemannienne compacte, les limites semiclassiques de la premi\`ere fonction propre associ\'ee \`a un opérateur positif du second ordre positif divers quand la
constante de diffusion $\epsilon$ tend vers z\'ero. Nous supposons que le terme d'ordre un est  un champ 
de vecteur $b$, dont les ensembles r\'ecurrents sont des points hyperboliques ou des cycles ou des tores à 
deux dimensions. Les limites de la fonction propre normalis\'ee sont concentr\'ees sur les ensembles
r\'ecurrents de dimension maximale o\`u la pression topologique est atteinte. Sur le cycles et les tores, 
les mesures limites sont absolument continues par rapport \`a la mesure de probabilit\`e invariante par $b$. Nous avons d\'etermin\'e ces limites en utilisant une analyse de type blow-up. 

%{\it Pour citer cet article~: D. Holcman, I. Kupka, C. R. Acad. Sci.
%Paris, Ser. I 340 (2005).}

\end{abstract}

% now the Version franÁaise abrÈgÈe, if it exists
%\selectlanguage{francais}
%\section*{Version fran\c{c}aise abr\'eg\'ee}
% Text of your Version franÁaise abrÈgÈe here.
% Note you do not need to repeat here equations that you use in the
% main text - for example 'voir (3)' is quite acceptable.

\selectlanguage{english}
% main text

\section{Statement of the  problem }\label{Introduction}

On a compact Riemannian manifold $(M,g)$, we study the semi-classical limits 
of the first eigenfunction of a positive second order operator 
\begin{eqnarray}
{L}_{\epsilon}=\varepsilon \Delta _{g} +\theta (b)+c, 
\end{eqnarray}
when the diffusion constant $\epsilon$ goes to zero. $\Delta _{g}$ is 
the Laplace-Beltrami operator and $\theta (b)$ is the Lie derivative in the 
direction $b \in TM$. The function $c$ is chosen so that $%
L_{\epsilon }$ is positive. Note that $L_{\epsilon }$  cannot be conjugated in general to a
self-adjoint operator by scalar multiplication. The Krein-Rutman theorem implies that the first
 eigenvalue $\lambda _{\epsilon }$ is real positive, simple and the associated eigenfunction
 $u_{\epsilon }$ is of a fixed sign (see \cite{Kifer90,Pinsky}).
The  $L_2$ normalized  positive eigenfunction $u_{\epsilon }$,($\int_{V_{n}}u_{\epsilon }^{2}dV_{g} =1$), 
 is solution of 
\begin{eqnarray}  \label{edpfdt}
\epsilon \Delta _{g}u_{\epsilon }+(b,\nabla u_{\epsilon })+cu_{\epsilon }
=\lambda _{\epsilon }u_{\epsilon } .  \nonumber
\end{eqnarray}
 Let us introduce the following notation, 
\begin{eqnarray}
\Omega= b-\nabla \mathcal{L}\, ,v_{\epsilon }=e^{-\frac{\mathcal{L}}{%
2\epsilon }}u_{\epsilon }  \label{gauge} \,\, ,  c_{\epsilon } =\epsilon (c+\frac{\Delta _{g}\mathcal{L}}{2})+\Psi _{ \mathcal{L}},  \,\, , 
\Psi _{\mathcal{L}} =\frac{1}{4}(||{\nabla \mathcal{L}||}^{2}+2(\nabla 
\mathcal{L},\Omega )),  \nonumber
\end{eqnarray}%
where the function $\mathcal{L} M:\rightarrow \rR$ belongs to a special class of Lyapunov functions, associated to the vector field $ b$, as described in \cite{HK1}.
Equation (\ref{edpfdt}) is transformed into            
%\begin{eqnarray} \label{vrai}
$L_{\epsilon }(v_{\epsilon })=\epsilon ^{2}\Delta _{g}v_{\epsilon }+\epsilon
(\Omega ,\nabla v_{\epsilon })+c_{\epsilon }v_{\epsilon }=\epsilon \lambda
_{\epsilon }v_{\epsilon },$
and we impose the normalization condition $\int_{V_{n}}v_{\epsilon }^{2}dV_{g}=1.$

In this situation, a fundamental problem is to study 1) the limits of $\lambda_{\epsilon }$ 
2) the limits of the measures $v_{\epsilon }^2dV_g$, as $\epsilon$ goes to zero.

\noindent  \textbf{History of the problem} 

The study of $\lambda _{\epsilon }$ has its origin in the work of Kolmogorov about the mean first passage time (MFPT) of a random particle to the boundary of a bounded domain. For domains in $\rR^n$, very interesting contributions to the first problem were made, in particular by Friedlin-Wentzell \cite{FW} and Devinatz-Friedman \cite{Devi1,Fried}. In the global situation (compact Riemannian manifold),  Kifer  in \cite{Kifer90} made a fundamental contribution to the study of the limit of $\lambda _{\epsilon }$: When the recurrent set of the drift $b$ is a finite union of hyperbolic components \cite{Robinson}, $\lambda _{\epsilon }$ has a unique  limit, which can be explicitly determined in terms of the topological pressure of the components.

\noindent\textbf{Limits of the first eigenfunction}

Problem 2) was much less studied than problem 1). However in the local situation, there are some interesting contributions in the paper \cite{Devi2} by Devinatz-Friedman. Otherwise very little was known. Here, we study this problem 2) in the spirit of  Kifer.  We consider drift $b$ such that the recurrent set is the union of a finite number of hyperbolic components, which are  either points, cycles or 2-dimensional torii. 

Let us mention that problem 2) is related to questions  studied in the unique ergodic conjecture \cite{Sarnak},
 where the limits of eigenfunctions are studied.
%%%%%%%%%%%%%%%%%%%%%%%%%%%%%%%%%%%%%%%%%%%%
\section{ Notation and assumptions}

We consider here vector fields $b$, related to the Morse-Smale
fields, which satisfy some strong compatibility conditions with respect to the
metric $g$. More specifically we shall assume that for each field $b$:

\textrm{I)}The recurrent set is a finite union of stationary points, limit
cycles and two dimensional torii.

\textrm{II)} The stationary points are hyperbolic and for each such $P$ the
stable and unstable manifolds belonging to $P$ are orthogonal at $P$ with
respect to the metric $g.$

\textrm{III)} Any limit cycle $S$ has a tubular neighborhood $\mathcal{T}_{S}$
provided with a covering map $\Phi :\mathbb{R}^{m-1}\times \mathbb{R}%
\longrightarrow \mathcal{T}_{S}$ having the following properties:

(a) for all $(x^{\prime },\theta )\in \mathbb{R}^{m-1}\times \mathbb{R},\Phi
^{-1}\circ \Phi (x^{\prime },\theta )=\{(x^{\prime },\theta +nT_{S})|n\in 
\mathbb{Z}\}$, $T_{S}$ minimal period of S. 

(b) at any point ($0$,$\theta )\in $ $\mathbb{R}^{m-1}\times \mathbb{R}$,\ $%
\left( \Phi \right) ^{\ast }g_{(0,\theta )}=\sum_{n=1}^{m-1}dx_{n}^{2}
+g^{m m }(\theta )d\theta ^{2}$,($\theta=x^m$).

(c) at any point ($0$,$\theta )\in \mathbb{R}^{m-1}\times \mathbb{R}$,
$
\left( \Phi \right) _{\ast }b=\frac{\partial }{\partial \theta }%
+\sum_{i,j=1}^{m-1}B_{ij}x_{j}\frac{\partial }{\partial x_{i}},$
up to term of order two in $x^{\prime }=(x_{1},..,x_{m-1}),$ canonical
coordinates on $\mathbb{R}^{m-1}.$

(d) the $\left( m-1\right) \times (m-1)$ matrix $B=$\{$B_{ij}|1\leq i,j\leq
m-1$\} is hyperbolic and its stable and instable spaces are orthogonal with
respect to the Euclidean metric $\sum_{n=1}^{m-1}x_{n}^{2}$ on $\mathbb{R}%
^{m-1}$. 

\textrm{IV)} Any 2-dimensional torus $R$ has a tubular neighborhood $\mathcal{T}_{R}$
provided with a diffeomorphism $\Phi :\mathbb{R}^{m-2}\times \mathbb{R}%
^{2}\longrightarrow T_{R}$ having the following properties:

(a) at any point ($0$,$\theta )\in $ $\mathbb{R}^{m-2}\times \mathbb{R}^{2}$, 
$\theta =(\theta _{1},\theta _{2})$, $\theta _{1},\theta _{2}$ cyclic
coordinates,\ 

$\, \, \, \, \left( \Phi ^{-1}\right) ^{\ast }g_{(0,\theta
)}=\sum_{n=1}^{m-2}dx_{n}^{2}+a(\theta )d\theta _{1}^{2}+2b(\theta )d\theta
_{1}d\theta _{2}+c(\theta )d\theta _{2}^{2}$

(b) at any point ($0$,$\theta )\in \mathbb{R}^{m-2}\times \mathbb{R}^{2}$ ,
$
\, \, \, \, \left( \Phi \right) _{\ast }\left( b\right) (0,\theta )=\ \ k_{1}%
\frac{\partial }{\partial \theta ^{1}}+k_{2} \frac{\partial }{\partial
\theta ^{2}}+\sum_{i,j=1}^{m-2}B_{ij}x_{j}\frac{\partial }{\partial x_{i}} \hbox{ where: }
$

(i) the $\left( m-2\right) \times (m-2)$ matrix $B=$\{$B_{ij}|1\leq i,j\leq
m-2$\} is hyperbolic and its stable and instable spaces are orthogonal with
respect to the Euclidean metric $\sum_{n=1}^{m-2}x_{n}^{2}$ on $\mathbb{R}%
^{m-2}$,

(ii) $k_1,k_2 \in \hbox{\bb R}$ and $\frac{k_1}{k_2} \in \hbox{\bb R}-%
\hbox{\bb Q}$. 
% the vector field $\ b_{1}(\theta )\frac{\partial }{\partial \theta _{1}}%
%+b_{2}(\theta )\frac{\partial }{\partial \theta _{2}}$ on $\mathbb{T}$ can
%be $C^{\infty }$ conjugated to a constant field $k_{1}\frac{\partial }{%
%\partial \theta _{1}}+k_{2}\frac{\partial }{\partial \theta _{2}}$ where the
%ratio $\frac{k_{1}}{k_{2}}\in \mathbb{R}-\mathbb{Q}$. In the remainder of
%the paper, $(\theta _{1},\theta _{2})$ will denote the cyclic coordinates on 
%$\mathbb{T}$ in which $b$can be written  $k_{1}\frac{\partial }{\partial
%\theta _{1}}+k_{2}\frac{\partial }{\partial \theta _{2}}.$
A torus with such a flow will be called an irrational torus.

(iii) Small divisor condition: There exist constants $C>0,\alpha>0$ such that for all $m_{1},m_{2} \in \zZ $, $ |m_{1}k_{1}+m_{2}k_{2}| \geq C(m_{1}^{2}+m_{2}^{2})^{\alpha }  \label{sd}$

%%%%%%%%%%%%%%%%%%%%%%%%%%%%%%%%%%%%%%%%%%%%%%%%%%%%%
\section{Statement of the main result}
%%%%%%%%%%%%%%%%%%%%%%%%%%%%%%%%%%%%%%%%%%%%%%%%%%%%%
The spirit of our result can be summarized as follow:

\emph{\textquotedblleft\ On a Riemannian manifold, for any choice of a
special Lyapunov function $\mathcal{L}$, vanishing at order 2 on the
recurrent sets of the field, the limits as }$\varepsilon $ \emph{tends to 0
of the normalized measures $e^{-\frac{\mathcal{L}}{2\epsilon }}u_{\epsilon
}^{2}dV_{g}$ are concentrated on the components of the recurrent sets which
are of maximal dimension and where the topological pressure is achieved.}

The construction of the function  $\mathcal{L}$ is detailed in \cite{HK1}.
\begin{theorem}\label{th1} 
On a compact Riemannian manifold $(M,g)$, let $b$ be a 
Morse-Smale vector field and $\mathcal{L}$ be a special Lyapunov function
for $b$. Consider the normalized positive eigenfunction $u_{\epsilon }>0$ of
the operator $L_{\epsilon }=\epsilon \Delta +\theta (b)+c$, associated to
the first eigenvalue $\lambda _{\epsilon }$.
\begin{enumerate}
\item The recurrent set $S$ of $b$ is a union of a finite set of stationary points $S^s$, 
a finite set of periodic orbits $S^p$ and a finite set of two dimensional irrational torii $S^t$.
The limit set of a  normalized measure $\frac{ u_{\epsilon }^{2}e^{-\mathcal{L}/\epsilon }dV_{g}}
{\int_{V_{n}}u_{\epsilon}^{2}e^{-\mathcal{L}/\epsilon }dV_{g}}$, is contained in the set of probability measure $\mu$ of the form   
\begin{eqnarray*}
\mu =\sum_{P\in S^s_{tp}}c_{P}\delta _{P}+\sum_{\Gamma \in
S^p_{tp}}a_{\Gamma }\delta _{\Gamma }+\sum_{\hbox{\bbbis T}\in S^t_{tp}}b_{%
\hbox{\bbbis T}}\delta_{\hbox{\bbbis T} }
\end{eqnarray*}
where $S^s_{tp}$ (resp.$S^p_{tp},S^t_{tp}$) is the subset of $S^s$ (resp.$S^p,S^t$), where the topological pressure is attained. $\delta _{P}$ is the Dirac measure at P.

For $\Gamma \in S^p$ and h $\in C(V )$, 
$ \delta _{\Gamma }(h)= \int_{0}^{T_{\Gamma }} f_\Gamma(\theta) h(\Gamma
(\theta )) d\theta$, where $\theta \in \mathbb{R\longrightarrow }\Gamma (\theta )\in V$ is a
solution of $b$ representing $\Gamma$.%and $b_0(\theta) = d\theta(b(\Gamma(\theta)))$ 
(see the notations for the precise definition of $\theta$). The periodic function
 $f_\Gamma$ is given by 
\begin{eqnarray*}
f_\Gamma(\theta) = \exp \{  -{\int_0^\theta{ c(\Gamma(s)) } ds} + \frac{\theta}{T_\Gamma}\int_0^{T_\Gamma} c(\Gamma(s)) ds \}
\end{eqnarray*}
and $T_\Gamma$ is the minimal period of $\Gamma$.

$\hbox{ For {\bbbis T}} \in S^t \hbox{ and }\,  h \in C(V ), \delta_{\hbox{\bbbis T} }(h) =\int_{\hbox{\bbbis T}}h(\theta _{1},\theta
_{2}) f_{\hbox{\bbbis T}}(\theta _{1},\theta _{2})dS_{\hbox{\bbbis T}},$
where $dS_{\hbox{\bbbis T}}$ is the unique probability measure on $\hbox{\bbbis T}$ invariant under the action of the field b and $f_{\hbox{\bbbis T}}$ is the unique solution of maximum 1, of the equation 
\begin{eqnarray*}
k_1 \frac{\partial f}{\partial \theta_1}+k_2 \frac{\partial f}{\partial
\theta_2} +cf = \mu_2 f \hbox{ where } \mu_2=\int_{\hbox{\bbbis T}} c dS_{\hbox{\bbbis T}}.
\end{eqnarray*}
\item The coefficients $c_{P}, a_{\Gamma}, b_{\hbox{\bbbis T}}$ obey the following rule:
If at least one coefficient $b_{\hbox{\bbbis T}}>0$, then for all cycle $a_{\Gamma }=0$ and
 all points $c_{P}=0$. If all coefficients  $b_{\hbox{\bbbis T}}=0$, and at least one coefficient $a_{\Gamma }>0$ then all $c_{P}=0$.
\end{enumerate}
\end{theorem}

\section{Remark on the proofs and future perspectives }

The basic techniques are the Blow-up and stochastic analysis and the theory of Ornstein-Uhlenbeck operator, but they are too involved to be presented here. As a buy product, we estimate the decay of the first eigenfunction near the recurrent sets.

We conjecture that under $L_p$ normalization, where p$>$1, the limit measures are similar to the one we obtained, except that the coefficients are different. To our knowledge, it is still unknown if the limit measure is unique or not. But this is a difficult problem as explained partially in \cite{Simon} and in more detail in \cite{HK1}.  
%When b is Hamiltonian in the KAM situation, we conjecture that the limit measures are concentrated on The study of the limit measures,

% etc, etc

% \appendix

% \section{}
% \label{}

% The Acknowledgements are an un-numbered section
%\section*{Acknowledgements}

\end{document}